\documentclass[aps,twocolumn,amsmath,amssymb,superscriptaddress,prl]{revtex4-1}

\usepackage{graphicx}
\usepackage{bm}
\usepackage{color} 
\usepackage{comment}

\begin{document}

\title{Current-induced torque originating from orbital current}

\author{Yuya Tazaki}
\affiliation{Department of Applied Physics and Physico-Informatics, Keio University, Yokohama 223-8522, Japan}

\author{Yuito Kageyama} 
\affiliation{Department of Applied Physics and Physico-Informatics, Keio University, Yokohama 223-8522, Japan}

\author{Hiroki Hayashi}
\affiliation{Department of Applied Physics and Physico-Informatics, Keio University, Yokohama 223-8522, Japan}

\author{Takashi Harumoto}
\affiliation{School of Materials and Chemical Technology, Tokyo Institute of Technology, Tokyo 152-8552, Japan}

\author{Tenghua Gao}
\affiliation{Department of Applied Physics and Physico-Informatics, Keio University, Yokohama 223-8522, Japan}

\author{Ji Shi}
\affiliation{School of Materials and Chemical Technology, Tokyo Institute of Technology, Tokyo 152-8552, Japan}

\author{Kazuya Ando\footnote{Correspondence and requests for materials should be addressed to ando@appi.keio.ac.jp}}
\email{ando@appi.keio.ac.jp}
\affiliation{Department of Applied Physics and Physico-Informatics, Keio University, Yokohama 223-8522, Japan}
\affiliation{Keio Institute of Pure and Applied Sciences (KiPAS), Keio University, Yokohama 223-8522, Japan}
\affiliation{Center for Spintronics Research Network (CSRN), Keio University, Yokohama 223-8522, Japan}

\maketitle

\textbf{
The electrical manipulation of magnetization by current-induced spin torques has given access to realize a plethora of ultralow power and fast spintronic devices such as non-volatile magnetic memories~\cite{miron2011perpendicular,liu2012spinScience}, spin-torque nano-oscillators~\cite{demidov2012magnetic}, and neuromorphic computing devices~\cite{torrejon2017neuromorphic}. Recent advances have led to the notion that relativistic spin-orbit coupling is an efficient source for current-induced torques, opening the field of spin-orbitronics~\cite{RevModPhys.87.1213,RevModPhys.91.035004}.  Despite the significant progress, however, the fundamental mechanism of magnetization manipulation\textemdash the requirement of spin currents in generating current-induced torques\textemdash has remained unchanged. 
Here, we demonstrate the generation of current-induced torques without the use of spin currents.
By measuring the current-induced torque for naturally-oxidized-Cu/ferromagnetic-metal bilayers, we observed an exceptionally high effective spin Hall conductivity at low temperatures despite the absence of strong spin-orbit coupling. Furthermore, we found that the direction of the torque depends on the choice of the ferromagnetic layer, which counters the conventional understanding of the current-induced torque. These unconventional features are best interpreted in terms of an orbital counterpart of the spin torque, an orbital torque, which arises from the orbital Rashba effect and orbital current. These findings will shed light on the underlying physics of current-induced magnetization manipulation, potentially altering the landscape of spin-orbitronics.
}

The interplay between charge and spin currents, or the charge-spin conversion, is at the heart of spin-orbitronics, which aims to explore novel phenomena arising from spin-orbit coupling (SOC)~\cite{RevModPhys.87.1213,RevModPhys.91.035004}. The charge-to-spin conversion arising from the spin Hall effect and inverse spin galvanic effect, or Rashba-Edelstein effect, has been studied extensively~\cite{RevModPhys.87.1213}. Of particular recent interest is the current-induced spin-orbit torque (SOT), which originates from the charge-to-spin conversion in ferromagnetic-metal/heavy-metal heterostructures~\cite{AndoPRL,miron2011perpendicular,liu2012spinScience}. The SOT enables the electrical manipulation of the magnetization in the heterostructure, offering an opportunity to develop spintronic devices based on the relativistic SOC~\cite{RevModPhys.91.035004}. Since the charge-to-spin conversion relies on the SOC, heavy metals, such as Ta, W, Ir, and Pt, play an essential role to realize efficient generation of the SOT~\cite{RevModPhys.87.1213}. In contrast to heavy metals, the SOT generation efficiency of light metals with weak SOC is known to be negligible. The archetypal material with weak SOC is Cu, whose SOT generation efficiency is two orders of magnitude lower than that of Pt~\cite{RevModPhys.87.1213}.

A recent study has shown that the generation efficiency of the current-induced torque in Cu/Ni$_{81}$Fe$_{19}$ heterostructures is enhanced by more than two orders of magnitude through the natural oxidation of the Cu layer~\cite{an2016spin}. The torque generation efficiency of the naturally oxidized Cu is comparable to that of Pt despite the absence of heavy elements. This finding reveals the important role of the oxidation effect in exploring the physics and technology of spin-orbitronics. However, the mechanism behind this dramatic phenomenon remains unclear.

In this work, we address the origin of the dramatic enhancement of the torque efficiency induced by the natural oxidation of Cu. By measuring the current-induced torque in naturally-oxidized-Cu/Ni$_{81}$Fe$_{19}$ bilayers, we demonstrate that the high torque efficiency of the naturally oxidized Cu is further enhanced by decreasing temperature. At low temperature, the effective spin Hall conductivity of the naturally oxidized Cu reaches $\sigma_\text{SH}^\text{eff}=4\times 10^4$ $\Omega^{-1}$cm$^{-1}$, which is more than an order of magnitude higher than $\sigma_\text{SH}^\text{eff}$ of heavy metals and most of topological insulators. We show that the current-induced torque with the exceptionally high efficiency does not originate from the SOC in the naturally oxidized Cu. Furthermore, we found that the direction of the torque acting on the magnetization is reversed by replacing the Ni$_{81}$Fe$_{19}$ with Fe, which provides evidence that spin currents are not responsible for the observed torque. These unconventional features are best interpreted in terms of an orbital current generated by the orbital Rashba-Edelstein effect, arising from the broken inversion symmetry of the naturally oxidized Cu.

\begin{figure}[tb]
\center\includegraphics[scale=1]{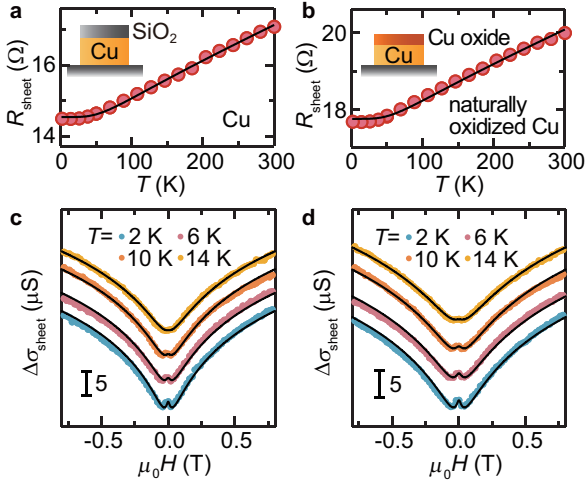}
\caption{
{\bfseries Weak antilocalization.} \textbf{a}, Temperature $T$ dependence of the sheet resistance $R_\mathrm{sheet}$ for the Cu film with the thickness of 10~nm. \textbf{b}, $T$ dependence of $R_\mathrm{sheet}$ for the naturally oxidized Cu film with the initial thickness of 10~nm. The red circles are the experimental data and the black curves are the result obtained using the Bloch-Gr{\"u}neisen formula. \textbf{c}, Magnetic field $H$ dependence of the magnetoconductance $\Delta \sigma_\text{sheet}(H) =\sigma_\text{sheet}(H)-\sigma_\text{sheet}({H = 0})$ for the Cu film. \textbf{d}, $H$ dependence of $\Delta \sigma_\text{sheet}(H)$ for the naturally oxidized Cu film. The solid circles are the experimental data. The black curves are the fitting result using the Hikami-Larkin-Nagaoka formula.
}
\label{WAL-NO} 
\end{figure}

First, to reveal the role of the SOC in the generation of the current-induced torque, we investigate a destructive interference between backscattered electron waves due to spin-orbit scattering: weak anti-localization (WAL), which allows to estimate the strength of the SOC~\cite{bergman1982influence}. We fabricated 10-nm-thick Cu films with and without a SiO$_2$ capping layer: Cu and naturally oxidized Cu films (see Methods and the inset to Figs.~\ref{WAL-NO}a and \ref{WAL-NO}b). In Figs.~\ref{WAL-NO}a and \ref{WAL-NO}b, we show temperature $T$ dependence of the sheet resistance $R_\mathrm{sheet}$ of these films. This result shows that the current flow in the oxidized layer is negligible in the naturally oxidized Cu. Here, a naturally oxidized Cu film is best described by a two-layer model, Cu-oxide/Cu (see the inset to Fig.~\ref{WAL-NO}b)~\cite{GATTINONI2015424,platzman2008oxidation,LIM20084040}. The Cu oxide layer consists of a duplex type oxide layer with an insulating CuO layer at the surface and an inner semiconducting Cu$_2$O layer: CuO/Cu$_2$O/Cu~\cite{GATTINONI2015424,an2016spin}. Since the electric resistivity of CuO and Cu$_2$O is orders of magnitude larger than that of Cu~\cite{DELOSSANTOSVALLADARES20126368}, almost all the applied charge current flows only in the Cu layer that has not been oxidized. The negligible current in the oxidized layer is supported by the fact that the temperature coefficient, $\alpha=(1/R_\mathrm{sheet})(dR_\mathrm{sheet}/dT)$, is almost unchanged by the natural oxidation (see Supplementary Information 1). In Figs.~\ref{WAL-NO}c and \ref{WAL-NO}d, we show the magnetoconductance $\Delta \sigma_\mathrm{sheet} (T, H) =  \sigma_\mathrm{sheet}(T, H) - \sigma_\mathrm{sheet}(T, H = 0)$, where $\sigma _\mathrm{sheet}(T, H)$ is the sheet conductance measured at the temperature $T$ and magnetic field $H$ applied perpendicular to the film. For both Cu and naturally oxidized Cu films, negative magnetoconductance was observed around $\mu_0 H=0$, which is attributed to the WAL effect. In fact, as shown in Figs.~\ref{WAL-NO}c and \ref{WAL-NO}d, the measured magnetoconductance $\Delta \sigma_\mathrm{sheet} (T, H)$ is well fitted by the Hikami-Larkin-Nagaoka formula~\cite{hikami1980spin,van1987resonant}:
\begin{multline}
\Delta\sigma_\text{WAL} (H)=-\frac{e^2}{2\pi^2\hbar}\left[
\psi \left(\frac{1}{2}+\frac{B_1}{\mu_0 H}\right) -\frac{3}{2}\psi \left(\frac{1}{2}+\frac{B_2}{\mu_0 H}\right)\right. \\ \left.+\frac{1}{2}\psi\left(\frac{1}{2}+\frac{B_\phi}{\mu_0 H}\right)-\ln\left( \frac{B_1\sqrt{B_\phi}}{B_2^{3/2}}\right)  
\right],\label{WALeq}
\end{multline}
where $\psi (x)$ is the digamma function. $B_1$, $B_2$, and $B_\phi$ are characteristic magnetic fields defined by $B_1=B_\text{e}+B_\text{SO}+B_\text{m}$, $B_2=B_\text{i}+(4/3)B_\text{SO}+(2/3)B_\text{m}$, and $B_\phi=B_\text{i}+2B_\text{m}$, where $B_\text{e}$, $B_\text{i}$, $B_\text{SO}$, and $B_\text{m}$ are the characteristic magnetic field for the elastic, inelastic, spin-orbit, and magnetic impurity scattering mechanisms, respectively.

Here, we note that the WAL effect is almost unchanged by the natural oxidation of the Cu film, as is clear from Figs.~\ref{WAL-NO}c and \ref{WAL-NO}d. We determined $B_\text{SO}$ from the WAL at different $T$ using $B_2=B_\text{i}+(4/3)B_\text{SO}+(2/3)B_\text{m}$ based on the fact that $B_\text{m}$ is independent of $T$, while $B_\text{i}$ depends on $T$ (see Supplementary Information 2). From the value of $B_\text{SO}$, we obtained the probability of the spin-orbit scattering process on the elastic scattering process, $\tau_\text{SO}^{-1}/\tau^{-1}$, which characterizes the strength of the SOC~\cite{abrikosov1962spin}. The result shows that $\tau_\text{SO}^{-1}/\tau^{-1}$ is almost unchanged by the natural oxidation: $\tau_\text{SO}^{-1}/\tau^{-1}=2.5\times 10^{-4}$ for the Cu film and $\tau_\text{SO}^{-1}/\tau^{-1}=3.5\times 10^{-4}$ for the naturally oxidized Cu film, which is consistent with literature~\cite{PhysRevB.84.161405}. Furthermore, this value is orders of magnitude smaller than $\tau_\text{SO}^{-1}/\tau^{-1}\sim 10^{-1}$ of Pt~\cite{oxfordspin}, showing the weak SOC in the naturally oxidized Cu with the initial thickness of 10~nm.

\begin{figure}[tb]
\center\includegraphics[scale=1]{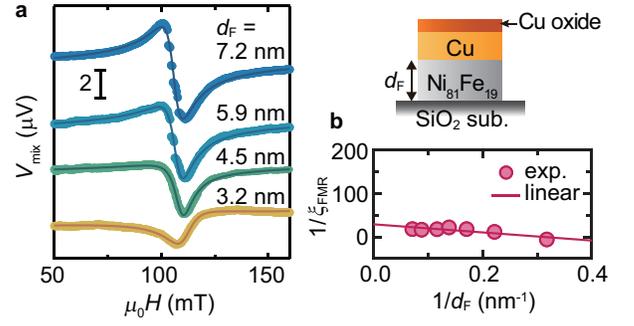}
\caption{
{\bfseries ST-FMR for naturally-oxidized-Cu/Ni$_{81}$Fe$_{19}$.} \textbf{a}, Magnetic field $H$ dependence of the DC voltage $V_\mathrm{mix}$ for the naturally-oxidized-Cu(10 nm)/Ni$_{81}$Fe$_{19}$($d_\text{F}$) bilayers with $d_\text{F}=3.2$ 4.5, 5.9, and 7.2 nm at 7 GHz. The solid circles are the experimental data and the solid curves are the fitting results using the sum of the symmetric and antisymmetric functions. \textbf{b}, 1/$d_\text{F}$ dependence of 
1/$\xi_\text{FMR}$ for the naturally-oxidized-Cu(10 nm)/Ni$_{81}$Fe$_{19}$($d_\text{F}$) bilayer. The solid circles are the experimental data. The solid lines are the linear fitting result. 
}
\label{thickness} 
\end{figure}

\begin{figure*}[tb]
\center\includegraphics[scale=1]{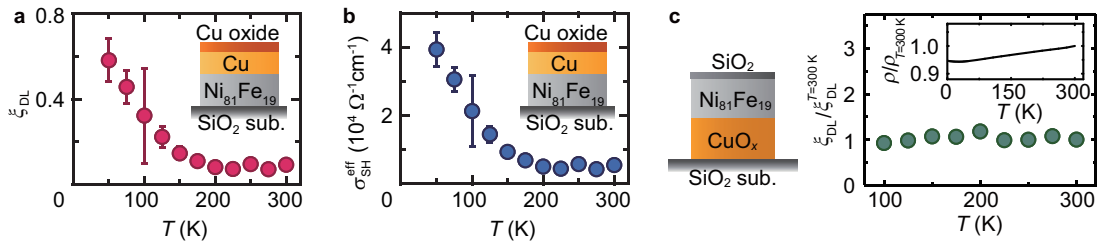}
\caption{
{\bfseries Temperature dependence of DL torque efficiency.} \textbf{a}, $T$ dependence of the DL-torque efficiency $\xi_\mathrm{DL}$ for the naturally-oxidized-Cu(10 nm)/Ni$_{81}$Fe$_{19}$ bilayer, where $\xi_\mathrm{DL}$ was determined by the 1/$d_\text{F}$ dependence of 1/$\xi_\text{FMR}$ at each $T$. The inset shows schematic illustration of the device. \textbf{b}, $T$ dependence of the effective spin Hall conductivity $\sigma_\text{SH}^\text{eff}$ for the naturally-oxidized-Cu/Ni$_{81}$Fe$_{19}$ bilayer. \textbf{c}, Schematic illustration of the Ni$_{81}$Fe$_{19}$/CuO$_x$ bilayer and temperature $T$ dependence of $\xi_\text{DL}$ for the Ni$_{81}$Fe$_{19}$/CuO$_x$(10 nm) bilayer, where $\xi_\text{DL}^{T=300\text{ K}} = 0.04$ is the DL spin-torque efficiency at $T=300$ K. $\xi_\text{DL}$ was obtained from $1/d_\text{F}$ dependence of $1/\xi_\text{FMR}$ at each $T$. The inset shows $T$ dependence of the resistivity $\rho$ of the CuO$_x$ film, where $\rho_{T=300\text{ K}}=8.4 \times 10^{-5}$ $\Omega$cm is the resistivity of the CuO$_x$ film at $T=300$ K. 
}
\label{temp} 
\end{figure*}

The WAL result demonstrates that the dramatic enhancement of the torque efficiency induced by the natural oxidation of Cu originates from unconventional mechanisms that do not require the SOC. 
Next, to explore the mechanism of the efficient torque generation without using the SOC, we investigate spin-torque ferromagnetic resonance (ST-FMR) for naturally-oxidized-Cu(10~nm)/Ni$_{81}$Fe$_{19}$($d_\mathrm{F}$) bilayers, where the numbers in parentheses represent the
thickness. For the ST-FMR measurements, an RF charge current was applied along the longitudinal direction of the device and an in-plane external magnetic field $H$ was applied with an angle of $45^\circ$ from the longitudinal direction of the device (see Methods). In the bilayer, the RF current generates the damping-like (DL) and field-like (FL) torques, as well as an Oersted field, driving magnetization precession in the adjacent ferromagnetic Ni$_{81}$Fe$_{19}$ layer~\cite{PhysRevB.92.064426}. The magnetization precession induces an oscillation of the bilayer resistance by the anisotropic magnetoresistance, resulting in the generation of a direct current (DC) voltage $V_\text{mix}$ through the mixing of the RF charge current and oscillating resistance~\cite{liu2011spin}: $V_\text{mix}=V_\text{sym}{W^2}/{[(\mu_0H-\mu_0H_\text{FMR})^2+W^2]}+V_\text{anti}{W(\mu_0H-\mu_0H_\text{FMR})}/{[(\mu_0H-\mu_0H_\text{FMR})^2+W^2]}$, 
where $W$ and $H_\text{FMR}$ are the spectral width and ferromagnetic resonance field, respectively. In the $V_\text{mix}$ signal, the magnitude of the symmetric component, $V_\text{sym}$, is proportional to the DL effective field $H_\text{DL}$, and the magnitude of the antisymmetric component, $V_\text{anti}$, is proportional to the sum of the Oersted field $H_\text{Oe}$ and FL effective field $H_\text{FL}$~\cite{PhysRevB.92.064426}. Since the electric voltage due to the spin pumping in this system is known to be negligible~\cite{PhysRevLett.122.217701}, we neglect this contribution in the following. We measured $V_\text{mix}$ using a bias tee.

In Fig.~\ref{thickness}a, we show the $V_\mathrm{mix}$ signals for the naturally-oxidized-Cu/Ni$_{81}$Fe$_{19}$($d_\mathrm{F}$) bilayers with different $d_\mathrm{F}$ measured at room temperature. This result shows that the sign of the antisymmetric component $V_\text{anti}$ is reversed by changing $d_\mathrm{F}$, which is consistent with the previous report~\cite{an2016spin}. The sign reversal of $V_\text{anti}$ arises from the competition between the $d_\mathrm{F}$-dependent FL effective field $H_\text{FL}$ and $d_\mathrm{F}$-independent Oersted field $H_\text{Oe}$. In the presence of the non-negligible $H_\text{FL}$, a so-called self-calibrated analysis based on the ratio $V_\text{sym}/V_\text{anti}$ cannot be used~\cite{PhysRevB.92.064426}. However, even in the presence of $H_\text{FL}$, the DL and FL-torque efficiencies can be determined by measuring the ST-FMR for devices with various $d_\mathrm{F}$ because of the different $d_\mathrm{F}$ dependence of $H_\text{FL}$ and $H_\text{Oe}$~\cite{PhysRevB.92.064426} (see Supplementary information 3). In Fig.~\ref{thickness}b, we show $1/d_\mathrm{F}$ dependence of $1/\xi_\mathrm{FMR}$, where $\xi_\text{FMR}=({V_\text{sym}}/{V_\text{anti}})({e\mu_{0}M_\text{s}d_\text{F} d_\text{N}}/{\hbar})\sqrt{1+{{M_\text{eff}}}/{H_\text{FMR}}}$ is the FMR spin-torque generation efficiency, obtained by fitting the measured ST-FMR spectra using the sum of the symmetric and antisymmetric functions. From the $1/d_\mathrm{F}$ dependence of $1/\xi_\mathrm{FMR}$, the DL(FL)-torque efficiencies, $\xi_\text{DL(FL)} = (2e/\hbar )\mu_{0}M_\text{s}d_\text{F}H_\text{DL(FL)}/j_\text{c}^\text{N}$, can be obtained using~\cite{PhysRevB.92.064426}
\begin{equation}
\frac{1}{\xi_\text{FMR}}=\frac{1}{\xi_\text{DL}} \left(1+\frac{\hbar}{e}\frac{\xi_\text{FL}}{\mu_0 M_\text{s}d_\text{F}d_\text{N}} \right), \label{thicknesseq}
\end{equation}
where $j_\text{c}^\text{N}$ and $d_\text{N}$ are the charge current density and the thickness of the naturally oxidized Cu layer where the RF current flows. We use $d_\mathrm{N}=d_\mathrm{Cu}=8.2$ nm because of the negligible current flow in the oxidized layer, where $d_\mathrm{Cu}$ is the thickness of the non-oxidized Cu layer (see Supplementary Information 1). $M_\text{s}$ and $M_\text{eff}$ are the saturation magnetization and effective demagnetization field of the Ni$_{81}$Fe$_{19}$ layer, respectively. From the data shown in Fig.~\ref{thickness}b, we obtained $\xi_\mathrm{DL}=0.03$ and $\xi_\mathrm{FL}=-0.02$. This result shows that $\xi_\mathrm{DL}$ and $\xi_\mathrm{FL}$ are enhanced significantly by the natural oxidation and comparable to those of Pt/Ni$_{81}$Fe$_{19}$ bilayers, consistent with the previous study~\cite{an2016spin}.

We found that the high DL-torque efficiency $\xi_\text{DL}$ is further enhanced by decreasing temperature $T$. To quantify the $T$ dependence of $\xi_\text{DL}$, we measured the ST-FMR for the naturally-oxidized-Cu/Ni$_{81}$Fe$_{19}$ bilayers with different $d_\mathrm{F}$ at various $T$. Figure~\ref{temp}a shows the $T$ dependence of $\xi_\text{DL}$ for the naturally-oxidized-Cu/Ni$_{81}$Fe$_{19}$ bilayer. This result shows that $\xi_\mathrm{DL}$ increases significantly with decreasing $T$, and $\xi_\text{DL}$ exceeds 50\% at $T=50$ K. Here, it is important to note that the significant $T$ dependence of $\xi_\mathrm{DL}$ is irrelevant to a change of the current-induced Oersted field induced by the change of $T$. The reason for this is that $\xi_\mathrm{DL}$ was determined from the $1/d_\mathrm{F}$ dependence of $1/\xi_\mathrm{FMR}$ measured at each $T$ using equation~(\ref{thicknesseq}), and thus the Oersted field contribution is eliminated from $\xi_\mathrm{DL}$ and $\xi_\mathrm{FL}$. From the obtained $\xi_\text{DL}(T)$, we estimated the effective spin Hall conductivity~\cite{PhysRevB.92.064426} $\sigma_\text{SH}^\text{eff}(T)=\sigma_\text{N}(T)\xi_\text{DL}(T)$, as shown in Fig.~\ref{temp}(c), where $\sigma_\text{N}(T)$ is the conductivity of the non-oxidized Cu layer. Figure~\ref{temp}(c) shows that the effective spin Hall conductivity of the naturally oxidized Cu reaches $\sigma_\text{SH}^\text{eff}=4\times 10^4$ $\Omega^{-1}$cm$^{-1}$ at $T=50$ K. This exceptionally high $\sigma_\text{SH}^\text{eff}$ is more than an order of magnitude higher than that of heavy metals and most of topological insulators~\cite{RevModPhys.87.1213}.

The significant enhancement of $\xi_\mathrm{DL}$ induced by decreasing $T$ is a unique feature of the naturally oxidized Cu. Recently, we studied the SOTs in Ni$_{81}$Fe$_{19}$/CuO$_x$ bilayers with various oxidation levels, where the uniformly oxidized CuO$_x$ layer was fabricated by the reactive sputtering~\cite{PhysRevLett.121.017202,Kageyamaeaax4278}. The SOT efficiency of the Ni$_{81}$Fe$_{19}$/CuO$_x$ bilayer becomes comparable to that of the naturally-oxidized-Cu/Ni$_{81}$Fe$_{19}$ bilayer only when the interface SOC is maximized by a fine tuning of the oxidation level~\cite{Kageyamaeaax4278}. Figure~\ref{temp}c shows the $T$ dependence of $\xi_\mathrm{DL}$ for the optimized Ni$_{81}$Fe$_{19}$/CuO$_x$ film (see also Methods). The $T$ dependence of $\xi_\mathrm{DL}$ was determined by measuring the ST-FMR for the Ni$_{81}$Fe$_{19}$/CuO$_x$ film with different $d_\mathrm{F}$; in the Ni$_{81}$Fe$_{19}$/CuO$_x$ bilayer, the applied current flows in the slightly oxidized, conducting CuO$_x$ layer, which allows to determine $\xi_\mathrm{DL}$ from the $1/d_\mathrm{F}$ dependence of $1/\xi_\mathrm{FMR}$ at each $T$ with equation~(\ref{thicknesseq}). Figure~\ref{temp}c shows that $\xi_\mathrm{DL}$ in the Ni$_{81}$Fe$_{19}$/CuO$_x$ bilayer is almost independent of $T$. Here, in the Ni$_{81}$Fe$_{19}$/CuO$_x$ bilayer, the SOT is dominated by the SOC at the Ni$_{81}$Fe$_{19}$/CuO$_x$ interface~\cite{Kageyamaeaax4278}. The clear difference in the $T$ dependence of $\xi_\mathrm{DL}$ supports that unconventional mechanisms are responsible for the current-induced torque in the naturally-oxidized-Cu/Ni$_{81}$Fe$_{19}$ film.

\begin{figure}[tb]
\center\includegraphics[scale=1]{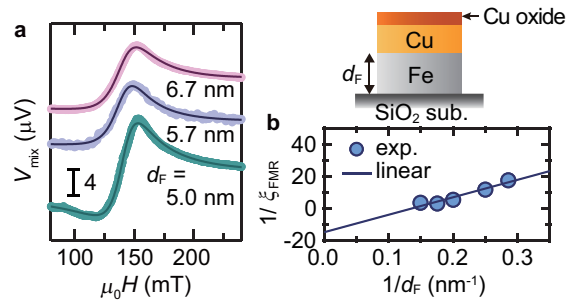}
\caption{
{\bfseries ST-FMR for naturally-oxidized-Cu/Fe.} \textbf{a}, Magnetic field $H$ dependence of the DC voltage $V_\mathrm{mix}$ for the naturally-oxidized-Cu(10 nm)/Fe($d_\text{F}$) bilayers at 13 GHz. The solid circles are the experimental data and the solid curves are the fitting results using the sum of the symmetric and antisymmetric functions. \textbf{b}, $1/d_\text{F}$ dependence of 1/$\xi_\text{FMR}$ for the naturally-oxidized-Cu/Fe bilayer. The solid circles are the experimental data. The solid lines are the linear fitting result. 
}
\label{STFMR-Fe} 
\end{figure}

\begin{table*}[tb]
  \begin{center}
    \caption{
        {\bfseries Classification of possible mechanisms of current-induced torque.} In the naturally-oxidized-Cu/ferromagnetic-metal (NO-Cu/FM) bilayer, several mechanisms can contribute to the observed torque. The origin of the SOC and charge current responsible for the possible torques are summarized. (i) The SOT due to the bulk spin Hall effect (SHE) can be induced by the SOC of the NO-Cu layer due to the current flow in the interior of the NO-Cu layer. Since a spin current is responsible for the torque, the sign of the DL torque is independent of the FM layer. The spin Rashba effect (SRE) at the (ii) Cu-oxide/Cu and (iii) Cu/FM interfaces can generate the torque in the NO-Cu/FM bilayer. The sign of the torque originating from the Cu/FM interface can depend on the FM layer, since the sign of the Rashba parameter can be different depending on the choice of the FM layer. In contrast, the sign of the SOT originating at the Cu-oxide/Cu interface is independent of the FM layer because the spin-transfer mechanism is responsible for the DL torque. (iv) The anomalous SOT can be induced by the SOC and current flow in the FM layer, and thus the sign depends on the FM layer. (v) The SOT can be generated by an inhomogeneous current flow in the NO-Cu layer through the spin-vorticity coupling (SVC). This process does not require the SOC. Since the torque arises from a spin current, the sign of the DL torque is independent of the FM layer. (vi) The orbital torque is generated by the current flow in the NO-Cu layer using the SOC in the FM layer. Since the SOC and band structure of the FM layer play an essential role, the sign of the orbital torque depends on the FM layer. The orbital torque is influenced by the natural oxidation, since the oxidation alters the orbital mixing that is responsible for the orbital Rashba effect. In these mechanisms, only the torque due to (i), (ii), (v) and (vi) can be changed by the natural oxidation, since the natural oxidation does not affect the Cu/FM interface (see Supplementary information 4). 
}\label{Classification}
\begin{tabular}{cccccc} \hline\hline
       & &\begin{tabular}{c}
SOC  \\[-3pt]
responsible for torque
\end{tabular}  & \begin{tabular}{c}
Charge current  \\[-3pt]
responsible for torque
\end{tabular}  & \begin{tabular}{c}
Sign of DL torque  \\[-3pt]
depends on FM
\end{tabular} & \begin{tabular}{c}
Effect of  \\[-3pt]
natural oxidation 
\end{tabular}   \\ \hline 
    (i) &SOT due to SHE & NO-Cu & NO-Cu & No & Possible   \\[-3pt]
    (ii)  &SOT due to SRE & Cu-oxide/Cu & Cu-oxide/Cu & No & Possible   \\[-3pt]
    (iii) &SOT due to SRE & Cu/FM  & Cu/FM & Yes & None \\[-3pt]
    (iv) &Anomalous SOT &  FM & FM &  Yes &  None  \\[-3pt]
    (v) &SOT due to SVC & None  & NO-Cu & No & Possible \\[-3pt]
    (vi)  &Orbital torque &  FM & NO-Cu &  Yes  & Possible   \\ [-3pt]
    (vii)  &Observed torque & FM or None  &  -  &  Yes  &   Remarkable  \\ 
      \hline\hline
    \end{tabular}
  \end{center}
\end{table*}

The origin of current-induced torques is usually attributed to spin currents created by the SOC. However, the current-induced DL-torque in the naturally-oxidized-Cu/Ni$_{81}$Fe$_{19}$ film cannot be attributed to spin currents. The evidence for this was obtained by replacing the Ni$_{81}$Fe$_{19}$ layer with Fe in the naturally-oxidized-Cu/Ni$_{81}$Fe$_{19}$ film. In Fig.~\ref{STFMR-Fe}a, we show the ST-FMR measured for naturally-oxidized-Cu(10 nm)/Fe($d_\mathrm{F}$) bilayers at room temperature. We plot the $1/d_\mathrm{F}$ dependence of $1/\xi_\mathrm{FMR}$, as shown in Fig.~\ref{STFMR-Fe}b. By fitting this result using equation~(\ref{thicknesseq}), we obtained $\xi_\mathrm{DL}=-0.07$ and $\xi_\mathrm{FL}=-0.15$ in the naturally-oxidized-Cu/Fe bilayer. What is notable is that the sign of the DL torque is reversed by replacing Ni$_{81}$Fe$_{19}$ with Fe: $\xi_\mathrm{DL}>0$ in the naturally-oxidized-Cu/Ni$_{81}$Fe$_{19}$ and $\xi_\mathrm{DL}<0$ for the naturally-oxidized-Cu/Fe. This sign reversal indicates that the DL torque cannot be attributed to spin currents; in the scenario where the current-induced torque originates from a spin current, the direction of the DL torque is reversed only when the spin-polarization direction of the spin current is reversed, and thus the direction of the DL torque is independent of the choice of the ferromagnetic layer~\cite{AndoPRL}.

The above unconventional features of the current-induced torque are best interpreted in terms of an orbital counterpart of the spin Rashba effect: the orbital Rashba effect~\cite{PhysRevLett.107.156803}. The orbital Rashba effect arises from the orbital mixing that is induced by breaking the spatial inversion symmetry~\cite{PhysRevLett.107.156803}. The orbital mixing induces Rashba-like chiral textures of the orbital angular momentum at the Fermi surface without using the SOC. In the naturally oxidized Cu, the spatial inversion symmetry is broken due to the oxidation of the surface. In the presence of the orbital Rashba effect in the naturally-oxidized-Cu/Ni$_{81}$Fe$_{19}$ bilayer, the orbital Rashba-Edelstein effect, induced by applying a charge current, generates an orbital current, which can give rise to a torque, an orbital torque, acting on the magnetization through the SOC of the ferromagnetic layer~\cite{PhysRevResearch.2.013177,arXiv:2002.00596}. This scenario does not require the SOC in the naturally oxidized Cu to generate a sizable current-induced torque, which is consistent with the observation of the coexistence of the high torque-generation efficiency and weak SOC of the naturally oxidized Cu. Furthermore, the most important feature of the orbital torque is that the sign and magnitude are determined by the SOC and band structure of the ferromagnetic layer, and therefore the sign of the DL torque can be reversed by changing the ferromagnetic layer~\cite{PhysRevResearch.2.013177}. A recent theoretical study predicts that the sign of the orbital torque is opposite between Ni and Fe~\cite{arXiv:2004.05945}. The observed sign reversal of the DL torque induced by replacing the ferromagnetic layer is consistent with this prediction. The increase of $\xi_\mathrm{DL}$ with decreasing $T$ is also consistent with this model because the orbital Rashba effect is suppressed by increasing $T$ due to a phonon-mediated electron hopping effect~\cite{chen2018giant}.

Finally, we summarize other possible sources of the current-induced torque. The possible mechanisms are summarized in Table~\ref{Classification}. In the naturally-oxidized-Cu/ferromagnetic-metal film, the observed torque cannot be attributed to the spin Hall effect and spin Rashba effect in the naturally oxidized Cu ((i) and (ii) in Table~\ref{Classification}) because of the weak SOC of the naturally oxidized Cu, evidenced by the WAL. These mechanisms are also inconsistent with the sign reversal of the DL torque induced by replacing the ferromagnetic layer. The spin Rashba effect at the Cu/ferromagnetic-metal interface ((iii) in Table~\ref{Classification}) is also not the source of the observed torque. Here, we confirmed that the natural oxidation does not affect the Cu/Ni$_{81}$Fe$_{19}$ interface in the naturally-oxidized-Cu/Ni$_{81}$Fe$_{19}$ bilayer (see Supplementary information 4). This indicates that the contribution from the Cu/Ni$_{81}$Fe$_{19}$ interface to the current-induced torque in the naturally-oxidized-Cu/Ni$_{81}$Fe$_{19}$ film is comparable to that in a SiO$_2$/Cu/Ni$_{81}$Fe$_{19}$ film, where the Cu layer is protected from the oxidation. However, the current-induced torque is vanishingly small in the SiO$_2$/Cu/Ni$_{81}$Fe$_{19}$ film (see Supplementary information 4), demonstrating the minor role of the Cu/Ni$_{81}$Fe$_{19}$ interface in the generation of the current-induced torque. The minor role of the Cu/Ni$_{81}$Fe$_{19}$ interface is also supported by the distinct difference in the $T$ dependence of $\xi_\mathrm{DL}$ between the naturally-oxidized-Cu/Ni$_{81}$Fe$_{19}$ and Ni$_{81}$Fe$_{19}$/CuO$_x$ films, where the origin of the SOT in the Ni$_{81}$Fe$_{19}$/CuO$_x$ film is the interfacial SOC.

The negligible effect of the natural oxidation on the Cu/Ni$_{81}$Fe$_{19}$ interface also shows that the observed torque cannot be attributed to the anomalous SOT, which is the torque generated by a ferromagnetic metal itself due to the asymmetry in the top and bottom interfaces~\cite{WangASOT} ((iv) in Table~\ref{Classification}). The negligible effect of the natural oxidation on the Cu/Ni$_{81}$Fe$_{19}$ interface indicates that the asymmetric feature of the Ni$_{81}$Fe$_{19}$ layer in the naturally-oxidized-Cu/Ni$_{81}$Fe$_{19}$ film is the same as that in the SiO$_2$/Cu/Ni$_{81}$Fe$_{19}$ film. Despite this fact, the sizable current-induced torque was observed only in the naturally-oxidized-Cu/Ni$_{81}$Fe$_{19}$ film, showing that the asymmetry in the Ni$_{81}$Fe$_{19}$ layer plays a minor role in the present systems. The negligible contribution from the anomalous SOT is also supported by the strong temperature $T$ dependence of the DL-torque efficiency, shown in Fig.~\ref{temp}a. The dramatic enhancement of $\xi_\mathrm{DL}$ induced by decreasing $T$ is clearly different from the $T$ dependence of the spin Hall and anomalous Hall effects, which are responsible for the anomalous SOT; the magnitude of the spin Hall and anomalous Hall effects at low temperature is comparable to that at room temperature in Ni$_{81}$Fe$_{19}$~\cite{PhysRevB.99.014403}. Although a recent study suggests that the SOT can also be generated by an inhomogeneous current flow in the naturally oxidized Cu through the spin-vorticity coupling~\cite{PhysRevLett.122.217701} ((v) in Table~\ref{Classification}), the observed DL torque is inconsistent with this model. The reason for this is that since the torque in this model is generated by a spin current, the direction of the DL torque is independent of the ferromagnetic layer (see also Supplementary Information 5). We also note that the opposite sign of the DL toque in the naturally-oxidized-Cu/Ni$_{81}$Fe$_{19}$ and naturally-oxidized-Cu/Fe films provides evidence that an unexpected Oersted field due to a possible inhomogeneity of the current flow in the naturally oxidized Cu is not the source of the current-induced torque, since the direction of the Oersted field is independent of the choice of the ferromagnetic layer. Our results therefore indicate that the observed current-induced torque can only be explained by the scenario where the torque is generated by the interior of the naturally oxidized Cu without using SOC, supporting the essential role of the orbital Rashba effect and orbital current.

In summary, we demonstrated the generation of the current-induced torque without the use of spin currents and SOC. The unique features of the observed torque are best interpreted in terms of an orbital counterpart of the spin torque, the orbital torque, which arises from the orbital Rashba effect and orbital current. The exceptionally high torque efficiency without the use of spin currents potentially alters the landscape of the physics and technology of spin-orbitronics. In particular, the exceptionally high effective spin Hall conductivity of the naturally oxidized Cu makes it a promising source of the current-induced torque, since high spin Hall conductivity, that is the coexistence of high electrical conductivity and DL-torque efficiency, is the essential factor for ultralow-power device operation.

\bigskip\noindent
\textbf{Methods}

\bigskip\noindent\textbf{Samples for the WAL measurement.} 
For the fabrication of the Cu film, a 10-nm-thick Cu film was sputtered on a thermally oxidized Si substrate, and then the 4-nm-thick SiO$_2$ was sputtered on the Cu surface without breaking the vacuum to prevent the natural oxidation of the Cu layer. For the fabrication of the naturally oxidized Cu film, a 10-nm-thick Cu film, sputtered on a thermally oxidized Si substrate, was exposed to the laboratory ambient for five hours. The Cu and naturally oxidized Cu films were patterned into Hall bars with the width of $510$ $\mu$m and length of $1150$ $\mu$m.

\bigskip\noindent\textbf{Samples for the ST-FMR measurement.} 
The naturally-oxidized-Cu/Ni$_{81}$Fe$_{19}$ films were fabricated on SiO$_2$ substrates using magnetron sputtering. The base pressure in the chamber before the deposition was better than 1$\times$10$^{-5}$ Pa. First, the Ni$_{81}$Fe$_{19}$ layer with the thickness $d_\mathrm{F}$ was sputtered on the SiO$_2$ substrate, and then the 10-nm-thick Cu layer was sputtered without breaking the vacuum (see Fig.~\ref{thickness}). The surface was exposed to the laboratory ambient for five hours. The Ni$_{81}$Fe$_{19}$($d_\mathrm{F}$)/CuO$_x$(10 nm) films, used for the temperature dependence measurement, were fabricated on SiO$_2$ substrates using magnetron sputtering. For the sputtering of the uniformly oxidized CuO$_x$ film, we introduced the mixture of argon and oxygen gas into the chamber with the oxygen to argon gas flow ratio $Q=1\%$. To avoid the oxidation of the surface of the Ni$_{81}$Fe$_{19}$ layer, we first deposited the CuO$_x$ layer on a SiO$_2$ substrate. After the CuO$_x$ deposition, the chamber was evacuated, and then the Ni$_{81}$Fe$_{19}$ layer was sputtered on top of the CuO$_x$ layer in an Ar atmosphere. Without breaking the vacuum, the surface of the Ni$_{81}$Fe$_{19}$ layer was covered by the SiO$_2$ capping layer, which was sputtered from a SiO$_2$ target in an Ar atmosphere (see Fig.~\ref{temp}). The films were patterned into rectangular strips with 20 $\mu$m width and 150 $\mu$m length using the photolithography and lift-off techniques.


\bigskip\noindent
Correspondence and requests for materials should be addressed to K.A. (ando@appi.keio.ac.jp)
\\

\bigskip\noindent
\textbf{Acknowledgements}\\
We acknowledge H. W. Lee for helpful discussions. This work was supported by JSPS KAKENHI Grant Numbers 19H00864, 26220604, the Canon Foundation, the Asahi Glass Foundation, and Spintronics Research Network of Japan (Spin-RNJ).

\bigskip\noindent
\textbf{Additional information}\\
The authors declare no competing financial interests.

\bigskip\noindent
\textbf{Author contributions}\\
Y.T., Y.K, and H.H. fabricated devices. Y.T., Y.K., H.H., T.H., T.G., and J.S. collected and analyzed the data. K.A. and Y.T. designed the experiments and developed the explanation. K.A. and Y.T. wrote the manuscript. All authors discussed results and reviewed the manuscript.

\end{document}